\documentclass [onecolumn]{kluwer}

\usepackage{graphicx}
\usepackage{amsbsy}

\begin{document}
\begin{article}
\begin{opening}

\title{Eccentricity generation in hierarchical triple systems with non-coplanar and initially circular orbits}

\author{Nikolaos \surname{Georgakarakos}\email{ng@maths.ed.ac.uk}}
\institute{School of Mathematics, Edinburgh
University, Mayfield Road, Edinburgh EH9 3JZ, UK}
\runningtitle{Eccentricity generation in HTS}
\runningauthor{Nikolaos Georgakarakos}

\begin{abstract}
In a previous paper, we developed a technique for estimating the inner
eccentricity in coplanar hierarchical triple systems on initially
circular orbits, with comparable masses and with well separated
components, based on an expansion of the rate of
change of the Runge-Lenz vector.  Now, the same technique is extended to
non-coplanar orbits.  However, it can only be applied to
systems with ${I_{0}<39.23^{\circ}}$ or ${I_{0}>140.77^{\circ}}$,
where ${I}$ is the inclination of the two orbits, because of
complications arising from the so-called 'Kozai effect'.  The
theoretical model is tested against results from numerical integrations of the
full equations of motion.
\end{abstract}

\end{opening}

\section{Introduction}

In a previous paper \cite{geo}, we derived a formula for estimating the eccentricity of the inner binary of a
hierarchical triple system with well separated components and initially circular orbits.  However,
the derivation was based on the assumption that the orbits of the
three bodies were on the same plane.  In the current paper, we extend
the derivation to the more general case of non-coplanar orbits.
Generally, the method is the same as in the coplanar regime, although
more terms are included in the derivation of the short
period equations to improve the accuracy of the model.  There are also some minor changes in the calculation of the secular
contribution to the eccentricity.  

Other recent work on the dynamics of
hierarchical triple system includes the work done by
\inlinecite{ford}, \inlinecite{kis} and \inlinecite{kry}.
\section{Theory}
The theoretical model is constructed in the same way as in the
coplanar case: we derive expressions for the short period terms by using the definition of
the Runge-Lenz vector, while the secular evolution is studied by
means of canonical perturbation theory.  The combination of the short
period and secular part of the eccentricity is achieved by considering
the eccentricity (inner or outer) to consist of a short period and a long period (secular) component,
i.e. ${e=e_{{\rm short}}+e_{{\rm sec}}}$ (one can picture this by
recalling the expansion of the disturbing function in solar system
dynamics, where the perturbing potential is given as a sum of an
infinite number of cosines of various frequencies).  Thus,
considering the eccentricity to be initially zero leads to ${e_{{\rm
short}}=-e_{{\rm sec}}}$ (initially), which implies that, although the
eccentricity is initially zero, the short period and secular eccentricity may
not be.
\subsection{Calculation of the short-period contribution to the eccentricity}	
\label{s1}
The equation of motion of the inner binary, using the Jacobi notation
(${\boldsymbol {r}}$ for the relative position vector of the inner
binary and ${\boldsymbol {R}}$ the vector from the centre of mass of
${m_{1}}$ and ${m_{2}}$ to the outer mass ${m_{3}}$), is:
\begin{equation}
\ddot{\boldsymbol {r}}=-G(m_{1}+m_{2})\frac{\boldsymbol{r}}{r^{3}}+\boldsymbol{F},
\label{eqmo}
\end{equation}
where ${\boldsymbol{F}}$, the perturbation to the inner binary motion, is  
\begin{eqnarray}
\boldsymbol{F} & = & Gm_{3}(\frac{\boldsymbol{R}-\mu_{1}\boldsymbol{r}}{|\boldsymbol{R}-\mu_{1}\boldsymbol{r}|^{3}}-\frac{\boldsymbol{R}+\mu_{2}\boldsymbol{r}}{|\boldsymbol{R}+\mu_{2}\boldsymbol{r}|^{3}})=Gm_{3}\frac{\partial}{\partial{\boldsymbol{r}}}(\frac{1}{\mu_{1}|\boldsymbol{R}-\mu_{1}\boldsymbol{r}|}+\nonumber\\
& & +\frac{1}{\mu_{2}|\boldsymbol{R}+\mu_{2}\boldsymbol{r}|})
\label{potenc1}
\end{eqnarray}
with
\begin{displaymath}
\mu_{{\rm i}}=\frac{m_{{\rm i}}}{m_{1}+m_{2}}, \hspace{0.5 cm} i=1,2.
\end{displaymath}
Now, since we are dealing with hierarchical triple systems with well
separated components, implying that 
\begin{math}
r/R 
\end{math}
is small, the inverse distances in equation (\ref{potenc1}) can be expressed as:
\begin{displaymath}
\frac{1}{|\boldsymbol{R}-\mu_{1}\boldsymbol{r}|}=\frac{1}{R}\sum^{\infty}_{n=o}
\left( \frac{\mu_{1}r}{R} \right) ^{n}P_{{\rm n}}(\cos{\theta})
\end{displaymath}
and
\begin{displaymath}
\frac{1}{|\boldsymbol{R}+\mu_{2}\boldsymbol{r}|}=\frac{1}{R}\sum^{\infty}_{n=o}
\left(- \frac{\mu_{2}r}{R} \right)^{n}P_{{\rm n}}(\cos{\theta}),
\end{displaymath}
where ${P_{n}}$ are the Legendre polynomials and ${\theta}$ is the
angle between the vectors ${\boldsymbol{r}}$ and ${\boldsymbol{R}}$.  Expanding to third
order, the perturbation becomes
\begin{eqnarray}
\boldsymbol{F} & = & Gm_{3}\frac{\partial}{\partial{\boldsymbol{r}}}\left(\frac{3}{2}\frac{(\boldsymbol{r}\cdot
\boldsymbol{R})^{2}}{R^{5}}-\frac{1}{2}\frac{r^{2}}{R^{3}}-\frac{5(\mu_{2}^{2}-\mu_{1}^{2})}{2}\frac{(\boldsymbol{r}\cdot\boldsymbol{R})^{3}}{R^{7}}+\right.\nonumber\\
& & \left.+\frac{3(\mu_{2}^{2}-\mu_{1}^{2})}{2}\frac{r^{2}(\boldsymbol{r}\cdot\boldsymbol{R})}{R^{5}}\right).
\label{fpert}
\end{eqnarray}
The first two terms in the above equation come from the quadrupole
term (${P_{2}}$), while the other two come from the octupole term
(${P_{3}}$).  

Using now the definition of the Runge-Lenz vector, we can obtain an expression for the inner eccentricity.  The
inner eccentric vector
${\boldsymbol{e}_{1}}$ is given by
\begin{equation}
\boldsymbol{e}_{1}=-\frac{\boldsymbol{r}}{r}+\frac{1}{\mu}(\dot{\boldsymbol{r}}\boldsymbol{\times}\boldsymbol{h}),
\label{ecve}
\end{equation}
where
\begin{math}
\boldsymbol{h}=\boldsymbol{r}\boldsymbol{\times}\dot{\boldsymbol{r}}
\end{math}
and
\begin{math}
\mu=G(m_{1}+m_{2}).
\end{math}
Assuming that ${\boldsymbol{r}\cdot\dot{\boldsymbol{r}}=0}$, i.e. the inner
binary remains nearly circular (bear in mind that all bodies are on initially circular orbits), differentiating equation (\ref{ecve})
and substituting for ${\boldsymbol{F}}$ we obtain:
\begin{eqnarray}
\dot{\boldsymbol{e}}_{1} & = & \frac{Gm_{3}}{\mu
R^{3}}\left[\left(6\frac{(\boldsymbol{r}\cdot\boldsymbol{R})(\dot{\boldsymbol{r}}\cdot\boldsymbol{R})}{R^{2}}-15(\mu_{2}^{2}-\mu_{1}^{2})\frac{(\boldsymbol{r}\cdot\boldsymbol{R})^{2}(\dot{\boldsymbol{r}}\cdot\boldsymbol{R})}{R^{4}}+\right.\right.\nonumber\\
& &
+\left.3(\mu_{2}^{2}-\mu_{1}^{2})\frac{r^{2}(\dot{\boldsymbol{r}}\cdot\boldsymbol{R})}{R^{2}}\right)\boldsymbol{r}+\left(r^{2}-3\frac{(\boldsymbol{r}\cdot\boldsymbol{R})^{2}}{R^{2}}+\frac{15}{2}(\mu_{2}^{2}-\right.\nonumber\\
& & \left.\left.-\mu_{1}^{2})\frac{(\boldsymbol{r}\cdot\boldsymbol{R})^{3}}{R^{4}}-\frac{9}{2}(\mu_{2}^{2}-\mu_{1}^{2})\frac{r^{2}(\boldsymbol{r}\cdot\boldsymbol{R})}{R^{2}}\right)\dot{\boldsymbol{r}}\right].
\end{eqnarray}
Now, choosing a frame of reference such that the initial plane of the
inner orbit is our reference plane and the line of nodes is initially
on the x-axis, with the positive direction of the x-axis pointing at
the ascending node of the outer orbit, the Jacobi vectors can be represented approximately in polar form as
\begin{math}
\boldsymbol{r}=a_{1}(\cos{n_{1}t},\sin{n_{1}t},0)
\end{math}
and
\begin{math}
\boldsymbol{R}=a_{2}(\cos{(n_{2}t+\phi)},\sin{(n_{2}t+\phi)}\cos{I},\sin{(n_{2}t+\phi)}\sin{I}),
\end{math}
where ${a_{1}}$ and ${a_{2}}$ are the semi-major axes of the inner and
outer orbit respectively and ${\phi}$ is the initial relative phase of the two binaries.
After integrating, the components ${x_{1}}$ and ${y_{1}}$ of the
eccentric vector become (expanding the coefficients in powers of
\begin{math}
\frac{1}{X}
\end{math}
and retaining the four leading terms, with ${X}$ being the period ratio
of the two orbits):
\begin{eqnarray}
x_{1} & = &
\frac{m_{3}}{M}\frac{1}{X^{2}}\left[P_{{\rm x21}}(t)+\frac{1}{X}P_{{\rm x22}}(t)+m_{*}\left(X^{\frac{1}{3}}P_{{\rm
x31}}(t)+\frac{1}{X^{\frac{2}{3}}}P_{{\rm x32}}(t)\right)\right]+\nonumber\\
& & +C_{{\rm x}_{1}}\label{e11}\\
y_{1} & = &
\frac{m_{3}}{M}\frac{1}{X^{2}}\left[P_{{\rm
y21}}(t)+\frac{1}{X}P_{{\rm y22}}(t)+m_{*}\left(X^{\frac{1}{3}}P_{{\rm
y31}}(t)+\frac{1}{X^{\frac{2}{3}}}P_{{\rm y32}}(t)\right)\right]+\nonumber\\
& & +C_{{\rm y}_{1}}\label{e12}
\end{eqnarray}
where ${P_{\rm i}(t)}$ are given
in the appendix and
\begin{equation}
m_{*}=\frac{m_{2}-m_{1}}{(m_{1}+m_{2})^{\frac{2}{3}}M^{\frac{1}{3}}}.
\end{equation}
${M}$ is the total mass of the system and ${C_{{\rm x}_{1}}}$ and
${C_{{\rm y}_{1}}}$ are
constants of integration.  The semi-major axes and mean motions were
treated as constants in the above calculation.

\subsection{Calculation of the secular contribution to the eccentricity}	
\label{s2}
Secular terms cannot be obtained
by the method of Sect. \ref{s1}, because, for an eccentric outer
binary, those terms appear as a linear function of time in the
expansion of the eccentric vector and therefore, they are valid for
limited time.  Thus, the secular contribution to the eccentricity
is studied by means of the Von Zeipel method.

The doubly averaged over short period terms for both orbits
Hamiltonian is (\opencite{mar}, with changes of notation):

\begin{eqnarray}
H & = &-\frac{Gm_{1}m_{2}}{2a_{{\rm
S}}}-\frac{G(m_{1}+m_{2})m_{3}}{2a_{{\rm T}}}+Q_{1}+Q_{2}, \label{hamilto} \\
\mbox{where}\nonumber\\
Q_{1} & = &
\frac{1}{8}\frac{Gm_{1}m_{2}m_{3}a^{2}_{{\rm
S}}}{(m_{1}+m_{2})a^{3}_{{\rm T}}(1-e^{2}_{{\rm
T}})^{\frac{3}{2}}}[-2-3e^{2}_{{\rm S}}+3\sin^{2}{I}(1-e^{2}_{{\rm
S}}+\nonumber\\
& & +5e^{2}_{{\rm S}}\sin^{2}{g_{{\rm S}}})],\\
Q_{2} & = &
\frac{15Gm_{1}m_{2}m_{3}(m_{2}-m_{1})a^{3}_{{\rm S}}e_{{\rm S}}e_{{\rm
T}}}{64(m_{1}+m_{2})^{2}a^{4}_{{\rm T}}(1-e^{2}_{{\rm
T}})^{\frac{5}{2}}}[(\sin{g_{{\rm S}}}\sin{g_{{\rm T}}}\cos{I}+\nonumber\\
& & +\cos{g_{{\rm S}}}\cos{g_{{\rm T}}})(4+3e^{2}_{{\rm
S}}-5\sin^{2}{I}(1-e^{2}_{{\rm S}}+7e^{2}_{{\rm S}}\sin^{2}{g_{{\rm
S}}}))-\nonumber\\
& & -10(1-e^{2}_{{\rm S}})\sin^{2}{I}\cos{I}\sin{g_{{\rm S}}}\sin{g_{{\rm T}}}].\label{q2}
\end{eqnarray}
The subscripts S and T denote the inner and outer longer period
orbit respectively.  The first term in the Hamiltonian is the Keplerian energy of the inner
binary, the second term is the Keplerian energy of the outer binary,
while the other two terms represent the interaction between the two
binaries.  The ${Q_{1}}$ term comes from the ${P_{2}}$ Legendre
polynomial and the ${Q_{2}}$ term comes from the ${P_{3}}$
Legendre polynomial.  There are also terms which arise from the
canonical transformation, but they are of smaller order than the
${P_{3}}$ term.  

By using Hamilton's equations (see \opencite{mar} for the derivation of
equations of motion involving the ${P_{2}}$ term), we can now derive the averaged
equations of motion of the system (see appendix), which are consistent
with the ones in \inlinecite{ford}, except that the ${P_{3}}$ term in
the Hamiltonian in \inlinecite{ford} has the wrong sign.  The same
sign error appears in \inlinecite{kry}.  After some exploratory numerical integrations of the five equations of
motion of the system, using a 4th-order Runge-Kutta method with
variable stepsize \cite{press}, it became clear that the outer
secular eccentricity and the inclination remained almost constant.  That, along
with the fact that the inner and outer eccentricities were not expected to reach
large values (which justifies neglecting powers of ${x_{{\rm S}}}$,
${y_{{\rm S}}}$ and ${e_{{\rm T}}}$ above the first order) and keeping
the dominant term in the equation for ${\dot{g}_{\rm T}}$ produced the following
simpler system of differential equations:

\begin{eqnarray}
\frac{{\rm d}g_{{\rm T}}}{{\rm d}\tau} & = & A \nonumber \\
\frac{{\rm d}x_{{\rm S}}}{{\rm d}\tau} & = & -By_{{\rm S}}+Ce_{{\rm T}}\sin{g_{{\rm T}}} \label{sys} \\
\frac{{\rm d}y_{{\rm S}}}{{\rm d}\tau} & = & Dx_{{\rm S}}-Ee_{{\rm T}}\cos{g_{{\rm T}}} \nonumber
\end{eqnarray}
where
\begin{displaymath}
x_{{\rm S}}=e_{{\rm S}}\cos{g_{{\rm S}}},\hspace{0.5cm}y_{{\rm S}}=e_{{\rm
S}}\sin{g_{{\rm S}}},
\end{displaymath}
\begin{displaymath}
A=\cos{I}+\frac{1}{2}\beta(4-5\sin^{2}{I}),\hspace{0.5cm}
B=2-5\sin^{2}{I}+\beta\cos{I},
\end{displaymath}
\begin{displaymath}
C=\frac{5}{16}\alpha \cos{I}(4-15\sin^{2}{I}),\hspace{0.1cm}
D=2+\beta\cos{I},\hspace{0.1cm}E=\frac{5}{16}\alpha (4-5\sin^{2}{I}),
\end{displaymath}
\begin{displaymath}
\alpha =\frac{m_{2}-m_{1}}{m_{1}+m_{2}}\frac{a_{{\rm S}}}{a_{{\rm T}}},\hspace{0.1cm}\beta
=\frac{m_{1}m_{2}M^{\frac{1}{2}}}{m_{3}(m_{1}+m_{2})^{\frac{3}{2}}}(\frac{a_{{\rm
S}}}{a_{{\rm T}}})^{\frac{1}{2}}\hspace{0.2cm}, {\rm d}\tau=\frac{3}{4}\frac{G^{\frac{1}{2}}m_{3}a^{\frac{3}{2}}_{{\rm
S}}}{a^{3}_{{\rm T}}(m_{1}+m_{2})^{\frac{1}{2}}}{\rm d}t.
\end{displaymath} 
The solution to system (\ref{sys}) is:
\begin{eqnarray}
g_{{\rm T}}(\tau) & = & A\tau+{g_{{\rm T}}}_{0}\\	
x_{{\rm S}}(\tau) & = &
K_{1}\cos{\sqrt{BD}\tau}+K_{2}\sin{\sqrt{BD}\tau}+\nonumber\\
& & +\frac{AC+BE}{BD-A^{2}}e_{{\rm T}}\cos{(A \tau+{g_{{\rm T}}}_{0})}\label{sacu1}\\
y_{{\rm S}}(\tau) & = &
K_{1}\sqrt{\frac{D}{B}}\sin{\sqrt{BD}\tau}-K_{2}\sqrt{\frac{D}{B}}\cos{\sqrt{BD}\tau}+\nonumber\\
& & +\frac{AE+CD}{BD-A^{2}}e_{{\rm T}}\sin{(A \tau+{g_{{\rm T}}}_{0})},\label{sacu2}
\end{eqnarray}
where
\begin{math}
K_{1}, K_{2}
\end{math}
are constants of integration and 
\begin{math}
{g_{{\rm T}}}_{0}
\end{math}
is the initial value of ${g_{{\rm T}}}$. It should be pointed out here
that although the initial outer eccentricity is zero, the secular one
is not and consequently an initial outer secular argument of
pericentre can be defined.  However, the above approximation to
secular motion is not valid if ${39.23^{\circ}<I<140.77^{\circ}}$, since the inner eccentricity is expected to
become significant due to the Kozai effect \cite{kozai} and therefore
the assumption that the eccentricity remains small is invalid in this case.

\subsubsection{Calculation of the initial outer secular eccentricity}
As was stated in the previous section, the outer secular
eccentricity remains almost constant.  Thus, the only thing that
remains now is to obtain an estimate for the initial outer secular
eccentricity.  This can be done in the following way:  
First, we find an expression for the short period outer eccentricity,
by following the same procedure as we did in Sect. (\ref{s1}), but this time we do it for the outer orbit.  The equation of
motion of the outer binary is 
\begin{equation}
\ddot{\boldsymbol{R}}=-GM\left(\mu_{1}\frac{\boldsymbol{R}+\mu_{2}\boldsymbol{r}}{|\boldsymbol{R}+\mu_{2}\boldsymbol{r}|^{3}}+\mu_{2}\frac{\boldsymbol{R}-\mu_{1}\boldsymbol{r}}{|\boldsymbol{R}-\mu_{1}\boldsymbol{r}|^{3}}\right)
\end{equation}
and eventually we obtain, to leading order, for the components of the outer short-period
eccentric vector:
\begin{eqnarray}
x_{2} & = &
\frac{M_{*}}{X^{\frac{4}{3}}}[\cos^{2}{I}(\frac{15}{16}\cos{(n_{2}t+\phi)}-\frac{7}{16}\cos{(3n_{2}t+3\phi)})+\nonumber\\
& & +\frac{7}{16}\cos{(3n_{2}t+3\phi)}-\frac{3}{16}\cos{(n_{2}t+\phi)}]+C_{x_{2}}\label{e21}\\
y_{2} & = &
\frac{M_{*}}{X^{\frac{4}{3}}}[\cos^{3}{I}(\frac{21}{16}\sin{(n_{2}t+\phi)}-\frac{7}{16}\sin{(3n_{2}t+3\phi)})+\nonumber\\
& & +\cos{I}(\frac{7}{16}\sin{(3n_{2}t+3\phi)}-\frac{9}{16}\sin{(n_{2}t+\phi)})]+C_{y_{2}}\label{e22}\\
z_{2} & = &
\frac{M_{*}}{X^{\frac{4}{3}}}[\cos^{2}{I}\sin{I}(\frac{21}{16}\sin{(n_{2}t+\phi)}-\frac{7}{16}\sin{(3n_{2}t+3\phi)})+\nonumber\\
& & +\sin{I}(\frac{7}{16}\sin{(3n_{2}t+3\phi)}-\frac{9}{16}\sin{(n_{2}t+\phi)})]+C_{z_{2}},\label{e23}\\
\mbox{where}
& & M_{*}=\frac{m_{1}m_{2}}{(m_{1}+m_{2})^{\frac{4}{3}}M^{\frac{2}{3}}}.
\end{eqnarray}
Suppose now that the outer secular eccentric vector is
${\boldsymbol{e}_{{\rm T}}=(x_{{\rm T}},y_{{\rm T}},z_{{\rm T}})}$.  Then, the constants
${C_{{\rm x}_{2}}}$, ${C_{{\rm y}_{2}}}$ and ${C_{{\rm z}_{2}}}$  in
equations (\ref{e21}), (\ref{e22}) and (\ref{e23}) can be replaced by
${x_{{\rm T}}}$, ${y_{{\rm T}}}$ and ${z_{{\rm T}}}$, since
${\frac{\delta x_{T}}{\delta x_{2}}<<1}$ (same for the pairs ${y_{T}-y_{2}}$ and ${z_{T}-z_{2}}$.  Considering that the outer binary is initially circular, i.e. ${e_{{\rm
out}}=0}$, the secular outer eccentric vector will initially be:
\begin{displaymath}
e_{{\rm T}_{0}}=(x_{{\rm T}_{0}},y_{{\rm T}_{0}},z_{{\rm T}_{0}})
\end{displaymath}
where
\begin{eqnarray}
x_{{\rm T}_{0}} & = &
-\frac{M_{*}}{X^{\frac{4}{3}}}[\cos^{2}{I}(\frac{15}{16}\cos{\phi}-\frac{7}{16}\cos{3\phi})+\frac{7}{16}\cos{3\phi}-\nonumber\\
& & -\frac{3}{16}\cos{\phi}]\\
y_{{\rm T}_{0}} & = &
-\frac{M_{*}}{X^{\frac{4}{3}}}[\cos^{3}{I}(\frac{21}{16}\sin{\phi}-\frac{7}{16}\sin{3\phi})+\cos{I}(\frac{7}{16}\sin{3\phi}\nonumber\\
& & -\frac{9}{16}\sin{\phi})]\\
z_{{\rm T}_{0}} & = &
-\frac{M_{*}}{X^{\frac{4}{3}}}[\cos^{2}{I}\sin{I}(\frac{21}{16}\sin{\phi}-\frac{7}{16}\sin{3\phi})+\sin{I}(\frac{7}{16}\sin{3\phi}-\nonumber\\
& & -\frac{9}{16}\sin{\phi})].\\
\end{eqnarray}
Thus, the initial outer secular eccentricity is:
\begin{equation}
e_{{\rm T}_{0}}=\sqrt{x^{2}_{{\rm T}_{0}}+y^{2}_{{\rm T}_{0}}+z^{2}_{{\rm T}_{0}}}.
\end{equation}

\subsection{A formula for the inner eccentricity}
\label{afft}
The expressions that were derived in paragraphs (\ref{s1}) and (\ref{s2}) for the
short period and secular contribution to the inner eccentric vector, can be combined to give an expression for the total eccentricity
in the same way we obtained the estimate for the initial outer secular eccentricity,
i.e. by replacing the constants in equations (\ref{e11}) and
(\ref{e12}) by equations (\ref{sacu1}) and (\ref{sacu2}), since
the latter evolve on a much larger timescale.  This yields (the minus
sign in front of ${x_{{\rm S}}}$ and ${y_{{\rm S}}}$ has to do with
the orientation of the coordinate system used in \opencite{mar}):
\begin{eqnarray}
x_{{\rm in}} & = & x_{1}-C_{{\rm x}_{1}}-x_{{\rm S}}\label{xtot}\\
y_{{\rm in}} & = & y_{1}-C_{{\rm y}_{1}}-y_{{\rm S}}\label{ytot}
\end{eqnarray}
The constants
${K_{1}}$ and ${K_{2}}$ in equations (\ref{sacu1}) and
(\ref{sacu2}) are determined by the fact that the inner eccentricity
is initially zero.

Averaging over time and over the initial relative phase ${\phi}$, the
averaged square inner eccentricity will be given by:
\begin{eqnarray}
\overline{e_{{\rm in}}^{2}} & = & <x^{2}_{{\rm in}}+y^{2}_{{\rm in}}>=\frac{m_{3}^{2}}{M^{2}}\frac{1}{X^{4}}\left[\frac{221}{64}-\frac{37}{32}\cos^{2}{I}+\frac{369}{64}\cos^{4}{I}+\right.\nonumber\\
& &
+\frac{D}{B}(\frac{19}{16}-\frac{5}{2}\cos^{2}{I}+3\cos^{4}{I})+\frac{B}{D}\cos^{2}{I}+\frac{1}{X^{2}}[\frac{1011}{144}+\nonumber\\
& &
+\frac{471}{24}\cos^{2}{I}+\frac{543}{144}\cos^{4}{I}+\frac{49}{9}\frac{D}{B}\cos^{2}{I}+\frac{B}{D}(\frac{121}{36}+\nonumber\\
& & +\frac{11}{9}\cos^{2}{I}+\frac{1}{9}\cos^{4}{I})]+\frac{1}{X}[\frac{45}{3}\cos{I}+\frac{93}{6}\cos^{3}{I}+\nonumber\\
& &
+\frac{D}{B}(\frac{7}{6}\cos{I}+\frac{14}{3}\cos^{3}{I})+\frac{B}{D}(\frac{11}{3}\cos{I}+\frac{2}{3}\cos^{3}{I})]+\nonumber\\
& &
+m^{2}_{*}X^{\frac{2}{3}}[\frac{1275}{8192}+\frac{36525}{8192}\cos^{2}{I}-\frac{103875}{8192}\cos^{4}{I}+\frac{76875}{8192}\cos^{6}{I}+\nonumber\\
& & +\frac{D}{B}(\frac{13925}{8192}\cos^{2}{I}-\frac{18875}{4096}\cos^{4}{I}+\frac{25625}{8192}\cos^{6}{I})+\frac{B}{D}(\frac{425}{8192}-\nonumber\\
& &
-\frac{875}{4096}\cos^{2}{I}+\frac{3125}{8192}\cos^{4}{I})]+\frac{m^{2}_{*}}{X^{\frac{4}{3}}}[\frac{138519}{131072}+\frac{62289}{131072}\cos^{2}{I}+\nonumber\\
& &
+\frac{121185}{131072}\cos^{4}{I}+\frac{102375}{131072}\cos^{6}{I}+\frac{D}{B}(\frac{54333}{131072}-\frac{42435}{65536}\cos^{2}{I}+\nonumber\\
& &
+\frac{119025}{131072}\cos^{4}{I})+\frac{B}{D}(\frac{94113}{131072}\cos^{2}{I}-\frac{17955}{65536}\cos^{4}{I}+\nonumber\\
& &
+\frac{10125}{131072}\cos^{6}{I})]+\frac{m^{2}_{*}}{X^{\frac{1}{3}}}[\frac{12495}{8192}\cos{I}-\frac{19875}{4096}\cos^{3}{I}+\nonumber\\
& &
+\frac{24375}{8192}\cos^{5}{I}+\frac{D}{B}(\frac{25545}{16384}\cos{I}-\frac{33975}{8192}\cos^{3}{I}+\frac{43125}{16384}\cos^{5}{I})+\nonumber\\
& &
+\left.\frac{B}{D}(-\frac{555}{16384}\cos{I}-\frac{5775}{8192}\cos^{3}{I}+\frac{5625}{16384}\cos^{5}{I})]\right]-\frac{m_{3}m_{*}M_{*}}{MX^{3}}\times\nonumber\\
& &
\times[(\frac{335}{1024}\cos{I}-\frac{875}{512}\cos{I}^{3}+\frac{1775}{1024}\cos{I}^{5})(1+\frac{D}{B})\frac{AC+BE}{BD-A^{2}}+\nonumber\\
& &
+(\frac{155}{1024}\cos{I}-\frac{335}{512}\cos{I}^{3}+\frac{875}{1024}\cos{I}^{5})(1+\frac{B}{D})\frac{AE+CD}{BD-A^{2}}]-\nonumber\\
& &
-\frac{m_{3}m_{*}M_{*}}{MX^{4}}[(\frac{219}{4096}-\frac{1935}{2048}\cos^{2}{I}+\frac{3795}{4096}\cos^{4}{I})(1+\frac{D}{B})\times\nonumber\\
& &
\times\frac{AC+BE}{BD-A^{2}}+(\frac{687}{4096}\cos^{2}{I}-\frac{1779}{2048}\cos^{4}{I}+\frac{1575}{4096}\cos^{6}{I})\times\nonumber\\
& &
\times(1+\frac{B}{D})\frac{AE+CD}{BD-A^{2}}]+\frac{M^{2}_{*}}{X^{\frac{8}{3}}}[\frac{(AC+BE)^{2}}{(BD-A^{2})^{2}}(1+\frac{D}{B})(\frac{29}{512}-\nonumber\\
& & -\frac{47}{256}\cos^{2}{I}+\frac{137}{512}\cos^{4}{I})+\frac{(AE+CD)^{2}}{(BD-A^{2})^{2}}(1+\frac{B}{D})\times\nonumber\\
& &
\times(\frac{65}{512}\cos^{2}{I}-\frac{119}{256}\cos^{4}{I}+\frac{245}{512}\cos^{6}{I})]+\frac{1}{2}\frac{M^{2}_{*}}{X^{\frac{8}{3}}}\times\nonumber\\
& &
\times\left[\frac{(AC+BE)^{2}}{(BD-A^{2})^{2}}+\frac{(AE+CD)^{2}}{(BD-A^{2})^{2}}\right](\frac{29}{256}-\frac{29}{256}\cos^{2}{I}-\nonumber\\
& & -\frac{101}{256}\cos^{4}{I}+\frac{245}{256}\cos^{6}{I}).
\label{final}
\end{eqnarray}

The above formula is expected to
be rather inaccurate in situations where the system parameters yield very small values for the
quantity ${BD-A^{2}}$, i.e. when we are near to a secular resonance.
Although the quantities ${\sqrt{BD}}$ and ${A}$ are approximations to
the secular frequencies of the inner and outer arguments of pericentre
respectively, the solution of ${BD-A^{2}=0}$, which reduces to
\begin{equation}
3(1-3\sin^{2}{I})(1-\beta^{2})-\frac{25}{4}\beta^{2}\sin^{4}{I}=0,
\end{equation}
could roughly identify
the location of the secular resonance.  However, in order to get a
more accurate solution to the problem, the inclusion of more terms is
necessary in system (\ref{sys}) or in the averaged Hamiltonian.

\section{Comparison with numerical results}
In order to test the range of applicability of the theory developed in the previous
sections, we integrated the full equations of motion numerically, using
a symplectic integrator with time transformation \cite{mik}.

The code calculates the relative position and velocity vectors of the
two binaries at every time step.  Then, by using standard two body formulae,
we computed the orbital elements of the two binaries.  Integrations were arranged such that the writing index ${Iwr}$ was ${1}$, the average number of steps per inner binary
period ${NS}$ was ${60}$, the method coefficients ${a1}$ and ${a2}$ were ${1}$ and ${15}$ respectively and
the correction index ${icor}$ was ${1}$.
In all simulations, we confined ourselves to systems with mass ratios within the
range ${10:1}$ since, ``among stellar triples, mass ratios are rare
outside a range of approximately ${10:1}$, although such systems would be
inherently difficult to recognise'' \cite{egg}; and
initial period ratio ${X \geq 10}$.  We also used units such that
${G=1}$ and ${m_{1}+m_{2}=1}$ and we always started the integrations
with ${a_{1}=1}$.  In that system of units, the initial conditions for the numerical
integrations were as follows:
\begin{displaymath}
r_{1}=1,\hspace{0.5cm} r_{2}=0,\hspace{0.5cm} r_{3}=0
\end{displaymath}	
\begin{displaymath}
R_{1}=a_{2}\cos{\phi},\hspace{0.5cm}
R_{2}=a_{2}\sin{\phi}\cos{I},\hspace{0.5cm} 	
R_{3}=a_{2}\sin{\phi}\sin{I}
\end{displaymath}	
\begin{displaymath}
\dot{r}_{1}=0,\hspace{0.5cm} \dot{r}_{2}=1,\hspace{0.5cm} \dot{r}_{3}=0
\end{displaymath}	
\begin{displaymath}
\dot{R}_{1}=-\sqrt{\frac{M}{a_{2}}}\sin{\phi},\hspace{0.5cm}
\dot{R}_{2}=\sqrt{\frac{M}{a_{2}}}\cos{\phi}\cos{I},\hspace{0.5cm}
\dot{R}_{3}=\sqrt{\frac{M}{a_{2}}}\cos{\phi}\sin{I},
\end{displaymath}	
where ${\boldsymbol {r}}$ and ${\boldsymbol{R}}$ are the relative
position vectors of the inner and outer orbit respectively.

\subsection{Short period effects} 
First we tested the validity of equations (\ref{e11}) and
(\ref{e12}).  The integrations and comparison with the analytical results were done for
${\phi=90^{\circ}}$, i.e. the outer binary was ahead of the inner one
at right angles.  However, this does not affect the qualitative
understanding of the problem at all. 
 
These results are presented in Table 1, which gives the percentage error between the
averaged, over time, numerical and theoretical ${e_{{\rm in}}}$ (the theoretical eccentricity was obtained by
evaluating  equations (\ref{e11}) and (\ref{e12}) everytime we
had an output from the symplectic integrator; both averaged
numerical and theoretical eccentricities  were calculated by using the
trapezium rule).  The
integrations were performed over one outer orbital period time span (in our
system of units, the initial outer orbital period is ${T_{{\rm out}}=2 \pi
X_{0}}$, where ${X_{0}}$ is the initial period ratio).
For each pair ${(m_{3},X_{0})}$ in Table 1, there are five entries,
corresponding, from top to bottom, to the following inner binaries:
${m_{1}=0.1-m_{2}=0.9}$, ${m_{1}=0.2-m_{2}=0.8}$,
${m_{1}=0.3-m_{2}=0.7}$, ${m_{1}=0.4-m_{2}=0.6}$ and
${m_{1}=0.5-m_{2}=0.5}$.  A dash in Table 1 denotes that the analogy among the masses was
outside the range ${10:1}$.  The inclination of the two orbits is ${I=20^{\circ}}$. 

\begin{figure}
\centerline{\includegraphics[width=10cm]{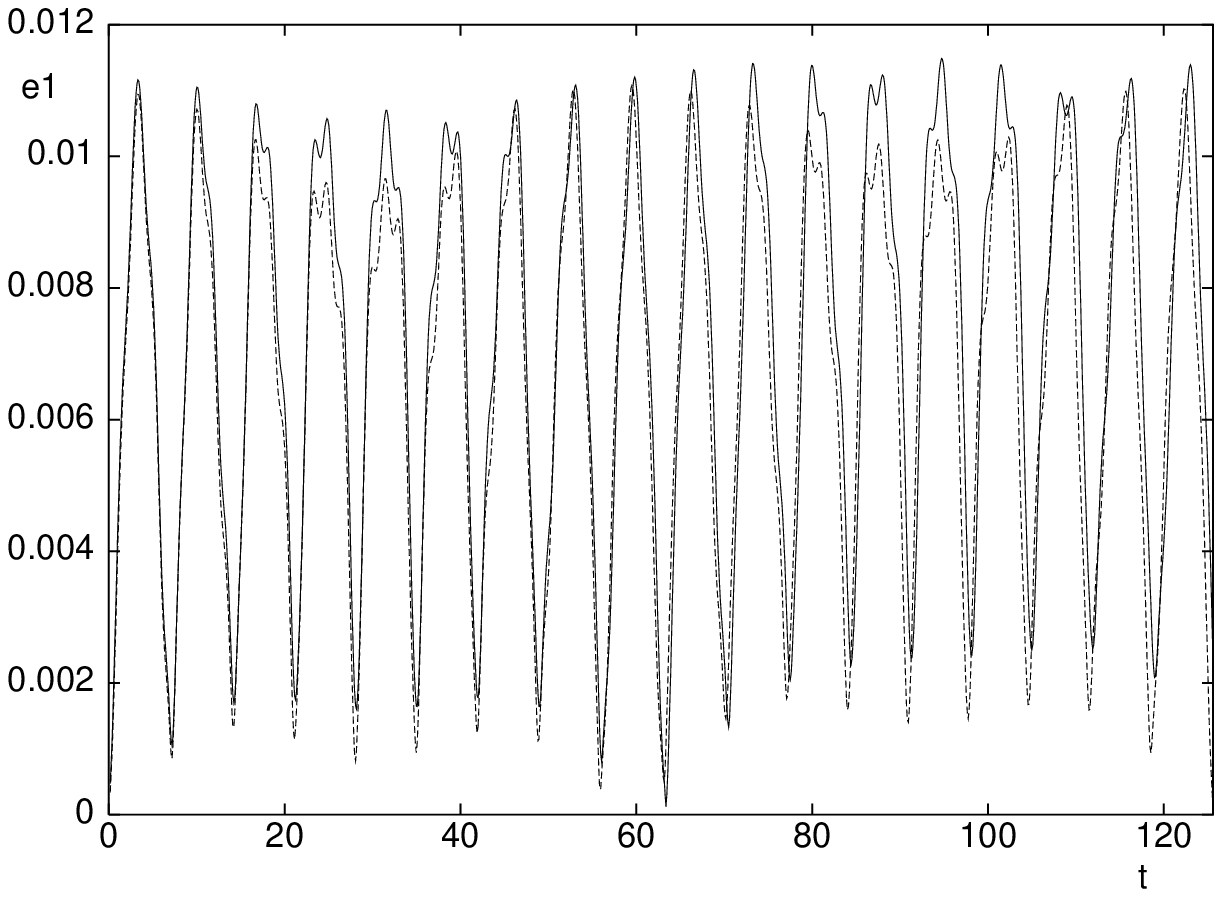}}
\caption{Inner eccentricity against time for a system with
${m_{1}=0.4}$, ${m_{3}=4}$, ${X_{0}=20}$, ${I=20^{\circ}}$ and ${\phi=90^{\circ}}$.  The
integration time span is one outer orbital period (${T_{{\rm out}}=125.6}$).  The continuous curve comes from 
the numerical integration of the full equations of motion, while the
dashed curve is a plot of equations (\ref{e11}) and (\ref{e12}).  In
the system of units used, the inner binary period is ${T_{{\rm in}}=2\pi}$.}
\label{fig0}
\end{figure}

Most of the results show a rather significant
error for systems with strong perturbation to the inner binary (small
${X_{0}}$-large ${m_{3}}$).  However, the error drops considerably as we move to
larger values of ${X_{0}}$ (the error becomes less than ${10 \%}$ for
all systems with ${X_{0}\geq 20}$).  This is consistent with our aim to obtain
a reasonable model for the evolution of the inner eccentricity in
hierarchical triple systems with well separated components.  One
should bear in mind that a period ratio of ${20}$ is close to the
lower boundary for observed hierarchical triple systems.  Fig.\ref{fig0}, which
is a plot of inner eccentricity against time, demonstrates the good
agreement between the theory (dashed curve) and the numerical results
(continuous curve).  The parameters of the system are: ${m_{1}=0.4}$,
${m_{3}=4}$, ${X_{0}=20}$, ${I=20^{\circ}}$, ${\phi=90^{\circ}}$ and the
integration time span is one outer orbital period (${T_{{\rm out}}=125.6}$).

\begin{table}
\caption[]{Percentage error between the
averaged numerical and averaged theoretical ${e_{{\rm in}}}$.  The
theoretical model is based on equations (\ref{e11}) and
(\ref{e12}).  For all systems, ${I=20^{\circ}}$ and ${\phi=90^{\circ}}$.  Each line corresponds to a different inner binary pair (${0.1-0.9, 0.2-0.8, 0.3-0.7, 0.4-0.6}$ and ${0.5-0.5}$).  A dash denotes that the mass ratio is outside the range ${10:1}$.}
\vspace{0.1 cm}
{\tiny \begin{tabular}{l l l l l l l l l l}\hline
${m_{3}\backslash\ X_{0}}$ & ${10}$ & ${15}$ & ${20}$ & ${25}$ & ${30}$ & ${50}$ \\
\hline
0.05 & - & - & - & - & - & - \\
     & - & - & - & - & - & - \\
     & - & - & - & - & - & - \\
     & - & - & - & - & - & - \\
     & 5.5 & 2.6 & 1.5 & 1 & 0.7 & 0.3 \\
0.09 & 5.1 & 2.5 & 1.6 & 1.1 & 0.9 & 0.4 \\
     & 5.6 & 2.7 & 1.7 & 1.2 & 0.9 & 0.4 \\
     & 5.9 & 2.8 & 1.7 & 1.2 & 0.9 & 0.4 \\
     & 6.1 & 2.9 & 1.8 & 1.2 & 0.9 & 0.4 \\
     & 6.2 & 3 & 1.8 & 1.2 & 0.9 & 0.4 \\
0.5  & 10 & 5.4 & 3.6 & 2.7 & 2.1 & 1.2 \\
     & 10.4 & 5.5 & 3.6 & 2.6 & 2.1 & 1.1 \\
     & 10.7 & 5.6 & 3.6 & 2.6 & 2 & 1 \\
     & 11 & 5.7 & 3.7 & 2.7 & 2.1 & 1 \\
     & 11.2 & 5.9 & 3.8 & 2.8 & 2.2 & 1.1 \\
 1   & 13.9 & 7.5 & 5 & 3.7 & 3 & 1.7 \\
     & 14.2 & 7.6 & 5 & 3.7 & 2.9 & 1.6  \\
     & 14.5 & 7.7 & 5 & 3.7 & 2.8 & 1.5  \\
     & 14.7 & 7.9 & 5.1 & 3.7 & 2.9 & 1.5  \\
     & 15 & 8.1  & 5.3 & 3.9 & 3 & 1.6 \\
1.5  & - & - & - & - & - & - \\
     & 16.7 & 8.9 & 5.9 & 4.3 & 3.4 & 1.8 \\
     & 16.9 & 9 & 5.9 & 4.3 & 3.4 & 1.8 \\
     & 17.3 & 9.3 & 6 & 4.4 & 3.4 & 1.8 \\
     & 17.7 & 9.5 & 6.3 & 4.6 & 3.6  & 1.9 \\
 2   & - & - & - & - & - & - \\
     & 18.5 & 9.9 & 6.5 & 4.8 & 3.7 & 2 \\
     & 18.7 & 10 & 6.5 & 4.8 & 3.7 & 1.9 \\
     & 19 & 10.2 & 6.7 & 4.9 & 3.8 & 2 \\
     & 19.4 & 10.4 & 6.9 & 5.1 & 4 & 2.1 \\
2.6  & - & - & - & - & - & - \\
     & - & - & - & - & - & - \\
     & 20.2 & 10.8 & 7 & 5.2 & 4 & 2.1 \\
     & 20.6 & 11 & 7.2 & 5.3 & 4.1 & 2.1 \\
     & 21 & 11.3 & 7.5 & 5.5 & 4.3 & 2.2 \\
 3   & - & - & - & - & - & - \\
     & - & - & - & - & - & - \\
     & 21 & 11.2 & 7.3 & 5.4 & 4.2 & 2.2 \\
     & 21.4 & 11.5 & 7.5 & 5.4 & 4.3 & 2.2 \\
     & 21.8 & 11.7 & 7.7  & 5.7  & 4.4 & 2.3 \\
3.4  & - & - & - & - & - & - \\
     & - & - & - & - & - & - \\
     & - & - & - & - & - & - \\
     & 22 & 11.8 & 7.7 & 5.6 & 4.4 & 2.3 \\
     & 22.4 & 12.1 & 8  & 5.8  & 4.6  & 2.4 \\
 4   & - & - & -  & - & - & - \\
     & - & - & - & - & - & - \\
     & - & - & - & - & - & - \\
     & 22.9 & 12.2 & 8 & 5.8 & 4.6 & 2.3 \\
     & 23.2 & 12.5 & 8.3  & 6 & 4.7 & 2.5 \\
4.5  & - & - & - & - & - & - \\
     & - & - & - & - & - & - \\
     & - & - & - & - & - & - \\
     & - & - & - & - & - & - \\
     & 23.8 & 12.8 & 8.4 & 6.2 & 4.8 & 2.5 \\
 5   & - & - & - & - & - & - \\
     & - & - & - & - & - & - \\
     & - & - & - & - & - & - \\
     & - & - & - & - & - & - \\
     & 24.2 & 13 & 8.6 & 6.3 & 4.9 & 2.6 \\
\hline
\end{tabular}}	
\end{table}
\subsection{Short and long period effects} 
Next, we tested equation (\ref{final}), which accounts for the short
period and secular effects to the inner eccentricity.  The formula was
compared with results obtained from integrating the full
equations of motion numerically.  These results are presented in Table 2, which
gives the percentage error between the averaged, over time and initial phase ${\phi}$, numerical
${e^{2}_{{\rm in}}}$ and equation (\ref{final}).  In the case of a system with noticeable secular evolution, the
error is accompanied by the period of the long oscillation
of the eccentricity, which is the same as the integration time span,
while the rest of the systems were
integrated over one outer orbital period (denoted by a dash), since there was not any
noticeable secular evolution when those systems were integrated over
longer time spans.  For each pair ${(m_{3}, X_{0})}$ in Table 2, there
are three pairs of entries, corresponding, from top to bottom, to
${10^{\circ}, 20^{\circ}}$ and ${30^{\circ}}$ inclination.  A negative
entry means that formula (\ref{final}) is an overestimate to the numerical
result.  Each system was numerically integrated for
${\phi=0^{\circ}-360^{\circ}}$ with a step of ${45^{\circ}}$.  After
the end of each simulation, ${e^{2}_{{\rm in}}}$ was averaged over time using the trapezium
rule and after the integrations for all ${\phi}$ were done, we
averaged over the relative initial phase by using the rectangle rule.  The integrations
were also done for smaller steps in ${\phi}$ (${10^{\circ}}$, ${1^{\circ}}$ and
${0.1^{\circ}}$), but there was not any difference in the outcome.
All the integrations presented in Table 2 were done for ${m_{1}=0.2}$
and ${m_{2}=0.8}$, but similar results are expected for the other inner
binary mass ratios.  

\begin{figure}
\centerline{\includegraphics[width=10cm]{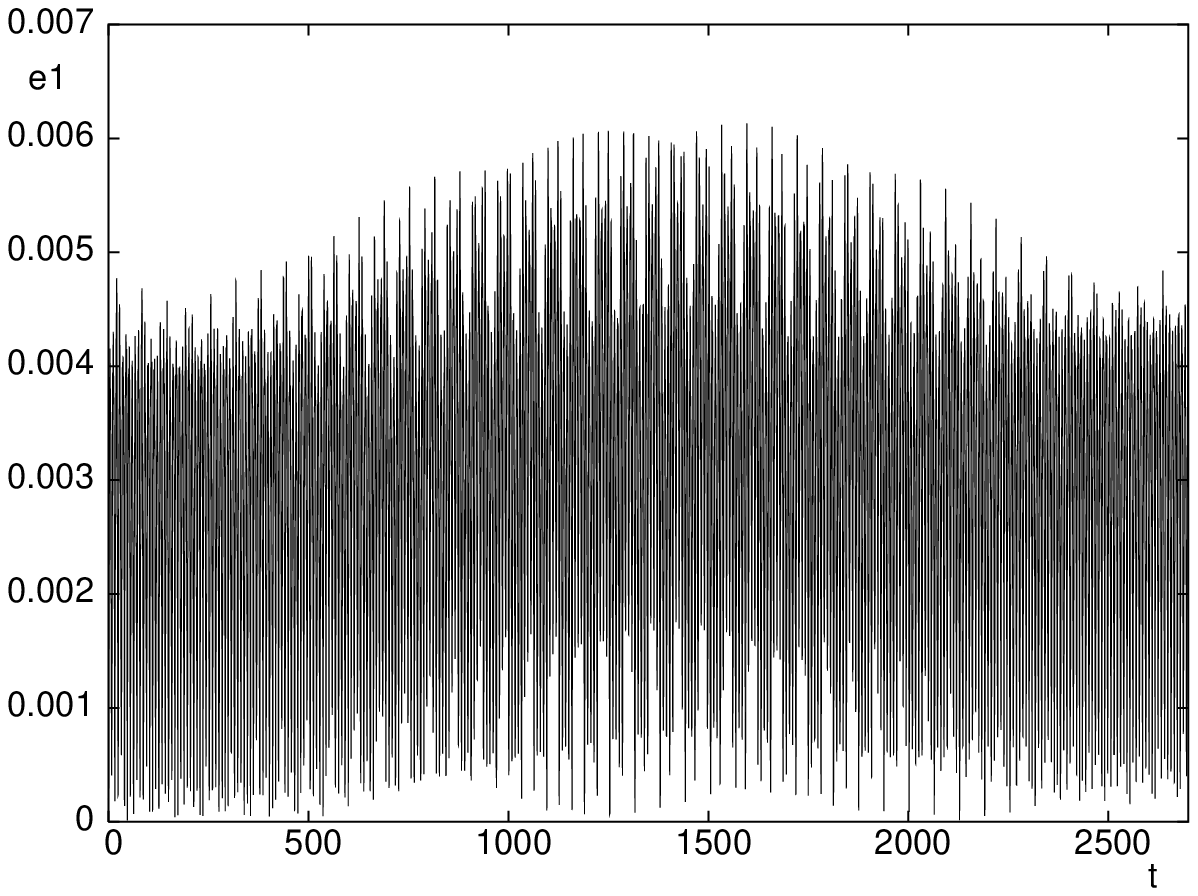}}
\centerline{\includegraphics[width=10cm]{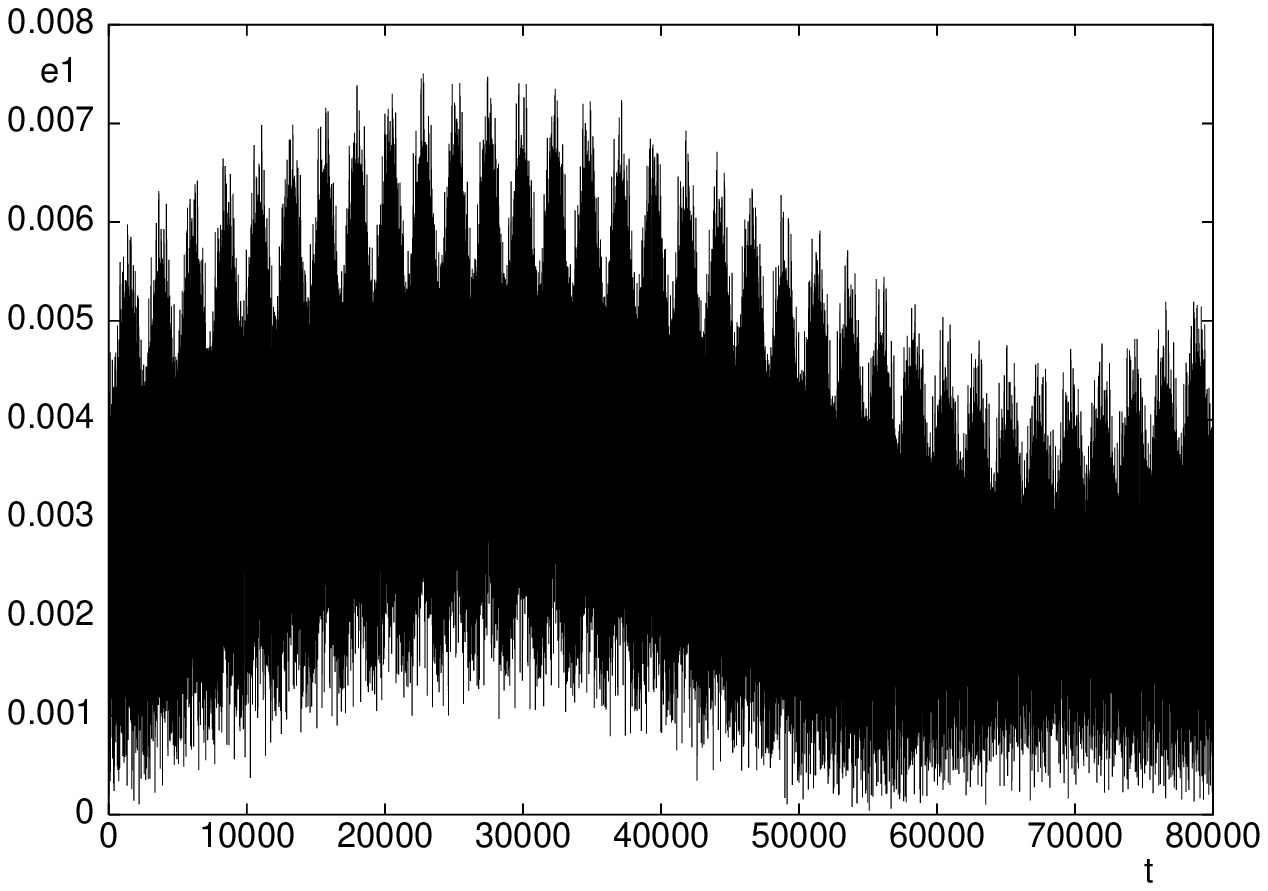}}
\caption{Secular resonance for a system with ${m_{1}=0.2}$,
${m_{3}=0.09}$, ${X_{0}=10}$, ${I=30^{\circ}}$ and
${\phi=90^{\circ}}$.  The outer orbital period is ${T_{{\rm
out}}=62.8}$.  The graphs come from the numerical
integration of the full equations of motion.  The top graph is a magnification of the first peak of the bottom graph.}
\label{fig1}
\end{figure}

\begin{figure}
\centerline{\includegraphics[width=10cm]{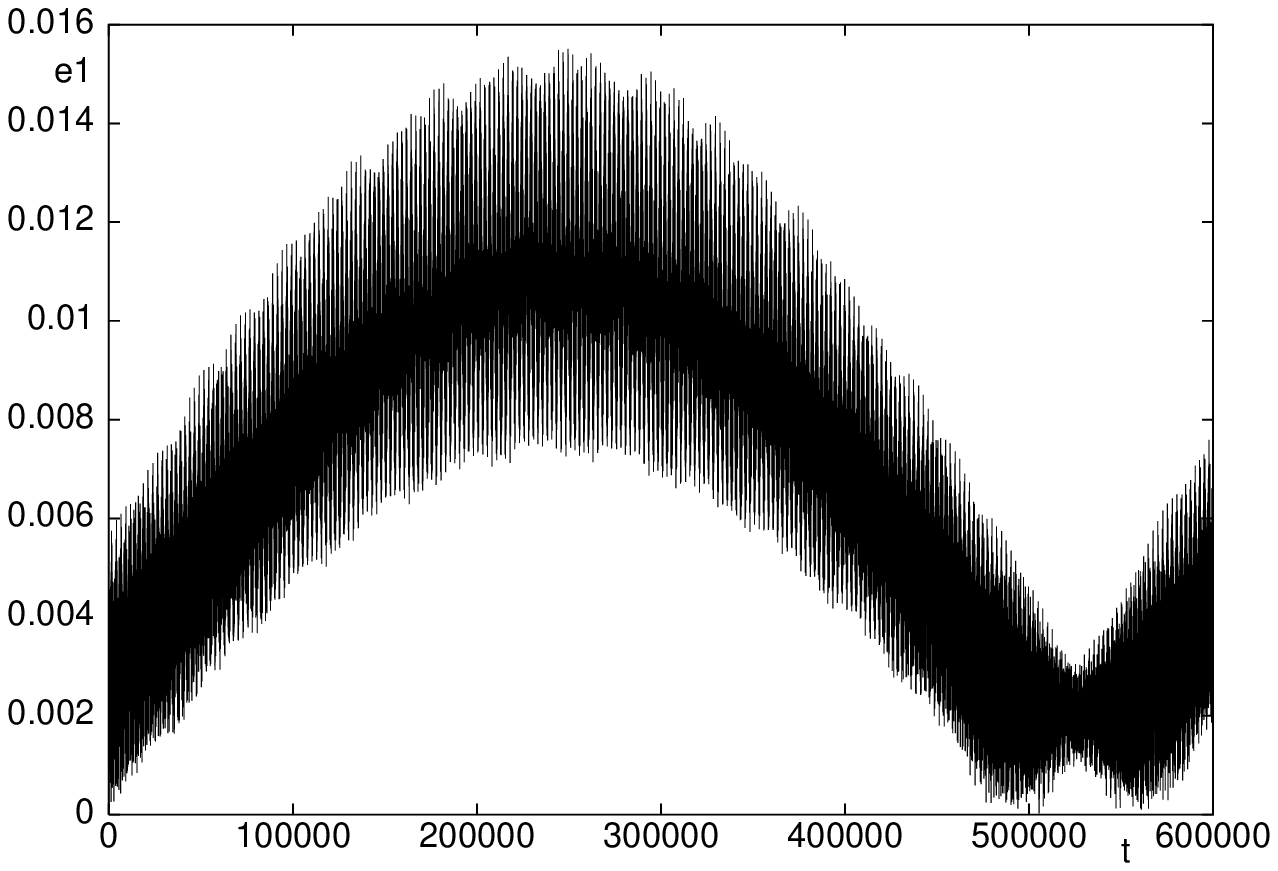}}
\caption{Secular resonance for a system with ${m_{1}=0.2}$,
${m_{3}=0.09}$, ${X_{0}=10}$, ${I=30^{\circ}}$ and ${\phi=90^{\circ}}$ based on
equations (\ref{xtot}) and (\ref{ytot}).  The outer orbital period is
${T_{{\rm out}}=62.8}$.  Note the long period and
large amplitude of the oscillation.}
\label{fig2}
\end{figure}

For systems with ${m_{3}=0.09}$, there is a significant discrepancy
between the numerical results and the theoretical model.  This is
clearly demonstrated in Figs. \ref{fig1} and \ref{fig2}.  It is easily noted in Fig. \ref{fig2},
which is a plot based on equations (\ref{xtot}) and (\ref{ytot}), that
the long period and amplitude of the oscillation are larger than the
ones obtained from the numerical integrations (Fig. \ref{fig1}).  This is due to the fact that the
system is in the vicinity of a secular resonance, i.e. the secular
frequencies of the two arguments of pericentre are nearly equal and
where, as stated in section(\ref{afft}), the secular part of our theory is not expected to
work very well.  The effect of the resonance gets less significant as we
move away from the solution of ${\sqrt{BD}-A=0}$.  However, for
systems with ${m_{1}=m_{2}}$ there is no such concern, since the long
term evolution of the system is independent of the outer argument of
pericentre ${g_{{\rm T}}}$, as can be seen in Sect. (\ref{s2}).

The rest of the numerical results are generally in good agreement with
our theory.  The error is just above or drops under ${10 \%}$ for all
masses and inclinations with ${X\geq 15}$.  Similar results are
expected for other inner mass ratios.

\begin{table}
\caption[]{Percentage error between the averaged numerical
${e^{2}_{{\rm in}}}$ and equation (\ref{final}) for systems with
${m_{1}=0.2}$ and ${m_{2}=0.8}$.  The error is accompanied by the integration time span.  A dash denotes that the system was integrated for one outer orbital period.  For each ${m_{3}-X_{0}}$ pair we have three entries corresponding, from top to bottom, to an inclination of ${10^{\circ}}$, ${20^{\circ}}$ and 
${30^{\circ}}$ respectively.}
\vspace{0.1 cm}
{\footnotesize \begin{tabular}{l l l l l l l l l l}\hline
${m_{3}\backslash\ X_{0}}$ & ${10}$ & ${15}$ & ${20}$ & ${25}$ & ${30}$ & ${50}$ \\
\hline
0.09 &-219.8&-51.3&-20.6&-12&-6.6&-0.7 \\
     &24000&62000&105000&160000&215000&510000\\
     &-245.1&-50.3&-21.5&-9.3&-5.9&0.3 \\
     &33500&81000&142000&200000&280000&650000\\
     &-1383.8&-396.3&-75.5&-33.9&-15.8&-2.6 \\
     &80000&210000&350000&550000&700000&1640000\\
0.5  &14.6&7.1&3.8&2.8&1.8&0.1 \\
     &-&-&-&3500&5200&15000\\
     &12.5&6.3&3.7&2.3&1.5&-0.1 \\
     &550&1400&2400&4100&6000&17000\\
     &8.2&2.8&1.5&0.7&0.1&-0.4 \\
     &650&1800&3200&5200&7800&22500\\
 1   &19.4&10.9&7.3&5.3&4.1&2.4 \\
     &-&-&-&-&-& 10000\\
     &14.8&8.2&5.5&4.2&3.2&1.6 \\
     &-&900&1800&2700&4200&12200\\
     &7.4&3.2&1.6&0.9&0.3&0.1 \\
     &450&1100&2100&3400&5200&15000 \\
1.5  &22&12.5&8.6&6.5&5.1&3 \\
     &-&-&-&-&-&9000 \\
     &16.8&9&6.4&4.9&3.9&2 \\
     &-&800&1400&2200&3400&9900\\
     &8.7&2.3&1&0.8&-0.1&-0.4 \\
     &300&900&1800&2800&4400&13000\\
 2   &23.7&13.4&9.2&7&5.7&3.1 \\
     &-&-&-&-&-&-\\
     &18.1&9.8&6.8&5.1&4.2&2.2 \\
     &-&600&1250&2000&3000&9000\\
     &8.1&2.6&0.8&0.7&0.1&0.1 \\
     &300&800&1600&2500&3800&11000\\
\hline
\end{tabular}}	
\end{table}


\section{Conclusion}

We have extended the method of getting an estimate for the inner
eccentricity in hierarchical triple systems on initially circular
orbits to situations where the stars are in non-coplanar orbits.
Again, the equations developed throughout this paper, give reasonable
results for the parameter ranges discussed.  The addition of two more
terms in the equations for short term evolution has made our
theoretical model more accurate, especially for systems where the
perturbation to the inner binary is rather strong (the equations in
the coplanar case only include the dominant ${P_{2}}$ and ${P_{3}}$
terms).  Our future aim is to complete that type of calculation by
deriving a formula for systems with eccentric outer binaries.
\appendix
\section{}
Short period components of the eccentric vector:
\begin{eqnarray}
P_{{\rm x21}}(t) & = &
\frac{11}{8} \cos{n_{1}t}+\frac{1}{8}\cos{3n_{1}t}+\frac{3}{16}\cos{((n_{1}-2n_{2})t-2\phi)}+\nonumber\\
& &
+\frac{3}{16}\cos{((n_{1}+2n_{2})t+2\phi)}+\frac{1}{16}\cos{((3n_{1}-2n_{2})t-2\phi)}+\nonumber\\
& &
+\frac{1}{16}\cos{((3n_{1}+2n_{2})t+2\phi)}+\cos{I}[\frac{9}{8}\cos{((n_{1}-2n_{2})t-2\phi)}-\nonumber\\
& &
-\frac{9}{8}\cos{((n_{1}+2n_{2})t+2\phi)}+\frac{1}{8}\cos{((3n_{1}-2n_{2})t-2\phi)}-\nonumber\\
& &
-\frac{1}{8}\cos{((3n_{1}+2n_{2})t+2\phi)}]+\cos^{2}{I}[-\frac{15}{8}\cos{n_{1}t}-\frac{1}{8}\cos{3n_{1}t}+\nonumber\\
& &
+\frac{15}{16}\cos{((n_{1}-2n_{2})t-2\phi)}+\frac{15}{16}\cos{((n_{1}+2n_{2})t+2\phi)}+\nonumber\\
& &
+\frac{1}{16}\cos{((3n_{1}-2n_{2})t-2\phi)}+\frac{1}{16}\cos{((3n_{1}+2n_{2})t+2\phi)}]\\
P_{{\rm x22}}(t) & = &
\frac{3}{8}\cos{((n_{1}-2n_{2})t-2\phi)}-\frac{3}{8}\cos{((n_{1}+2n_{2})t+2\phi)}+\nonumber\\
& &
+\frac{1}{24}\cos{((3n_{1}-2n_{2})t-2\phi)}-\frac{1}{24}\cos{((3n_{1}+2n_{2})t+2\phi)}+\nonumber\\
& &
+\cos{I}[\frac{9}{4}\cos{((n_{1}-2n_{2})t-2\phi)}+\frac{9}{4}\cos{((n_{1}+2n_{2})t+2\phi)}+\nonumber\\
& &
+\frac{1}{12}\cos{((3n_{1}-2n_{2})t-2\phi)}+\frac{1}{12}\cos{((3n_{1}+2n_{2})t+2\phi)}]+\nonumber\\
& &
+\cos^{2}{I}[\frac{15}{8}\cos{((n_{1}-2n_{2})t-2\phi)}-\frac{15}{8}\cos{((n_{1}+2n_{2})t+2\phi)}+\nonumber\\
& & +\frac{1}{24}\cos{((3n_{1}-2n_{2})t-2\phi)}-\frac{1}{24}\cos{((3n_{1}+2n_{2})t+2\phi)}]\\
P_{{\rm x31}}(t) & = &
\cos{I}[\frac{25}{64}\cos{(3n_{2}t+3\phi)}-\frac{165}{64}\cos{(n_{2}t+\phi)}]+\nonumber\\
& &
+\cos^{3}{I}[\frac{225}{64}\cos{(n_{2}t+\phi)}-\frac{25}{64}\cos{(3n_{2}+3\phi)}]\\
P_{{\rm x32}}(t)& = &
-\frac{69}{128}\cos{((2n_{1}-n_{2})t-\phi)}-\frac{69}{128}\cos{((2n_{1}+n_{2})t+\phi)}-\nonumber\\
& &
-\frac{15}{128}\cos{((2n_{1}-3n_{2})t-3\phi)}-\frac{15}{128}\cos{((2n_{1}+3n_{2})t+3\phi)}-\nonumber\\
& &
-\frac{45}{512}\cos{((4n_{1}-n_{2})t-\phi)}-\frac{45}{512}\cos{((4n_{1}+n_{2})t+\phi)}-\nonumber\\
& &
-\frac{15}{512}\cos{((4n_{1}-3n_{2})t-3\phi)}-\frac{15}{512}\cos{((4n_{1}+3n_{2})t+3\phi)}+\nonumber\\
& &
+\cos{I}[-\frac{21}{32}\cos{((2n_{1}-n_{2})t-\phi)}+\frac{21}{32}\cos{((2n_{1}+n_{2})t+\phi)}-\nonumber\\
& &
-\frac{15}{32}\cos{((2n_{1}-3n_{2})t-3\phi)}+\frac{15}{32}\cos{((2n_{1}+3n_{2})t+3\phi)}-\nonumber\\
& &
-\frac{45}{512}\cos{((4n_{1}-n_{2})t-\phi)}+\frac{45}{512}\cos{((4n_{1}+n_{2})t+\phi)}-\nonumber\\
& &
-\frac{45}{512}\cos{((4n_{1}-3n_{2})t-3\phi)}+\frac{45}{512}\cos{((4n_{1}+3n_{2})t+3\phi)}]+\nonumber\\
& &
+\cos^{2}{I}[\frac{75}{128}\cos{((2n_{1}-n_{2})t-\phi)}+\frac{75}{128}\cos{((2n_{}+n_{2})t+\phi)}-\nonumber\\
& &
-\frac{75}{128}\cos{((2n_{1}-3n_{2})t-3\phi)}-\frac{75}{128}\cos{((2n_{1}+3n_{2})t+3\phi)}+\nonumber\\
& &
+\frac{45}{512}\cos{((4n_{1}-n_{2})t-\phi)}+\frac{45}{512}\cos{((4n_{1}+n_{2})t+\phi)}-\nonumber\\
& &
-\frac{45}{512}\cos{((4n_{1}-3n_{2})t-3\phi)}-\frac{45}{512}\cos{((4n_{}+3n_{2})t+3\phi)}]+\nonumber\\
& &
+\cos^{3}{I}[\frac{45}{64}\cos{((2n_{1}-n2)t-\phi)}-\frac{45}{64}\cos{((2n_{1}+n_{2})t+\phi)}-\nonumber\\
& &
-\frac{15}{64}\cos{((2n_{1}-3n_{2})t-3\phi)}+\frac{15}{64}\cos{((2n_{1}+3n_{2})t+3\phi)}+\nonumber\\
& &
+\frac{45}{512}\cos{((4n_{1}-n_{2})t-\phi)}-\frac{45}{512}\cos{((4n_{1}+n_{2})t+\phi)}-\nonumber\\
& & -\frac{15}{512}\cos{((4n_{1}-3n_{2})t-3\phi)}+\frac{15}{512}\cos{((4n_{1}+3n_{2})t+3\phi)}]\\
P_{{\rm y21}}(t)& = &
-\frac{7}{8}\sin{n_{1}t}+\frac{1}{8}\sin{3n_{1}t}-\frac{15}{16}\sin{((n_{1}-2n_{2})t-2\phi)}-\nonumber\\
& &
-\frac{15}{16}\sin{((n_{1}+2n_{2})t+2\phi)}+\frac{1}{16}\sin{((3n_{1}-2n_{2})t-2\phi)}+\nonumber\\
& &
+\frac{1}{16}\sin{((3n_{1}+2n_{2})t+2\phi)}+\cos{I}[-\frac{9}{8}\sin{((n_{1}-2n_{2})t-2\phi)}+\nonumber\\
& &
+\frac{9}{8}\sin{((n_{1}+2n_{2})t+2\phi)}+\frac{1}{8}\sin{((3n_{1}-2n_{2})t-2\phi)}-\nonumber\\
& &
-\frac{1}{8}\sin{((3n_{1}+2n_{2})t+2\phi)}]+\cos^{2}{I}[\frac{3}{8}\sin{n_{1}t}-\frac{1}{8}\sin{3n_{1}t}-\nonumber\\
& &
-\frac{3}{16}\sin{((n_{1}-2n_{2})t-2\phi)}-\frac{3}{16}\sin{((n_{1}+2n_{2})t+2\phi)}+\nonumber\\
& &
+\frac{1}{16}\sin{((3n_{1}-2n_{2})t-2\phi)}+\frac{1}{16}\sin{((3n_{1}+2n_{2})t+2\phi)}]\\
P_{{\rm y22}}(t) &= &
-\frac{15}{8}\sin{((n_{1}-2n_{2})t-2\phi)}+\frac{15}{8}\sin{((n_{1}+2n_{2})t+2\phi)}+\nonumber\\
& &
+\frac{1}{24}\sin{((3n_{1}-2n_{2})t-2\phi)}-\frac{1}{24}\sin{((3n_{1}+2n_{2})t+2\phi)}+\nonumber\\
& &
+\cos{I}[-\frac{9}{4}\sin{((n_{1}-2n_{2})t-2\phi)}-\frac{9}{4}\sin{((n_{1}+2n_{2})t+2\phi)}+\nonumber\\
& &
+\frac{1}{12}\sin{((3n_{1}-2n_{2})t-2\phi)}+\frac{1}{12}\sin{((3n_{1}+2n_{2})t+2\phi)}]+\nonumber\\
& &
+\cos^{2}{I}[-\frac{3}{8}\sin{((n_{1}-2n_{2})t-2\phi)}+\frac{3}{8}\sin{((n_{1}+2n_{2})t+2\phi)}+\nonumber\\
& & +\frac{1}{24}\sin{((3n_{1}-2n_{2})t-2\phi)}-\frac{1}{24}\sin{((3n_{1}+2n_{2})t+2\phi)}]\\
P_{{\rm y31}}(t) & = &
\frac{25}{64}\sin{(3n_{2}t+3\phi)}-\frac{15}{64}\sin{(n_{2}t+\phi)}+\nonumber\\
& & +\cos^{2}{I}[\frac{75}{64}\sin{(n_{2}t+\phi)}-\frac{25}{64}\sin{(3n_{2}+3\phi)}]\\
P_{{\rm y32}}(t) & = &
\frac{33}{64}\sin{((2n_{1}-n_{2})t-\phi)}+\frac{33}{64}\sin{((2n_{1}+n_{2})t+\phi)}+\nonumber\\
& &
+\frac{15}{64}\sin{((2n_{1}-3n_{2})t-3\phi)}+\frac{15}{64}\sin{((2n_{1}+3n_{2})t+3\phi)}-\nonumber\\
& &
-\frac{45}{512}\sin{((4n_{1}-n_{2})t-\phi)}-\frac{45}{512}\sin{((4n_{1}+n_{2})t+\phi)}-\nonumber\\
& &
-\frac{15}{512}\sin{((4n_{1}-3n_{2})t-3\phi)}-\frac{15}{512}\sin{((4n_{1}+3n_{2})t+3\phi)}+\nonumber\\
& &
+\cos{I}[\frac{51}{128}\sin{((2n_{1}-n_{2})t-\phi)}-\frac{51}{128}\sin{((2n_{1}+n_{2})t+\phi)}+\nonumber\\
& &
+\frac{75}{128}\sin{((2n_{1}-3n_{2})t-3\phi)}-\frac{75}{128}\sin{((2n_{1}+3n_{2})t+3\phi)}-\nonumber\\
& &
-\frac{45}{512}\sin{((4n_{1}-n_{2})t-\phi)}+\frac{45}{512}\sin{((4n_{1}+n_{2})t+\phi)}-\nonumber\\
& &
-\frac{45}{512}\sin{((4n_{1}-3n_{2})t-3\phi)}+\frac{45}{512}\sin{((4n_{1}+3n_{2})t+3\phi)}]+\nonumber\\
& &
+\cos^{2}{I}[-\frac{15}{32}\sin{((2n_{1}-n_{2})t-\phi)}-\frac{15}{32}\sin{((2n_{1}+n_{2})t+\phi)}+\nonumber\\
& &
+\frac{15}{32}\sin{((2n_{1}-3n_{2})t-3\phi)}+\frac{15}{32}\sin{((2n_{1}+3n_{2})t+3\phi)}+\nonumber\\
& &
+\frac{45}{512}\sin{((4n_{1}-n_{2})t-\phi)}+\frac{45}{512}\sin{((4n_{1}+n_{2})t+\phi)}-\nonumber\\
& &
-\frac{45}{512}\sin{((4n_{1}-3n_{2})t-3\phi)}-\frac{45}{512}\sin{((4n_{1}+3n_{2})t+3\phi)}]+\nonumber\\
& &
+\cos^{3}{I}[-\frac{45}{128}\sin{((2n_{1}-n_{2})t-\phi)}+\frac{45}{128}\sin{((2n_{1}+n_{2})t+\phi)}+\nonumber\\
& &
+\frac{15}{128}\sin{((2n_{1}-3n_{2})t-3\phi)}-\frac{15}{128}\sin{((2n_{1}+3n_{2})t+3\phi)}+\nonumber\\
& &
+\frac{45}{512}\sin{((4n_{1}-n_{2})t-\phi)}-\frac{45}{512}\sin{((4n_{1}+n_{2})t+\phi)}-\nonumber\\
& & -\frac{15}{512}\sin{((4n_{1}-3n_{2})t-3\phi)}+\frac{15}{512}\sin{((4n_{1}+3n_{2})t+3\phi)}]
\end{eqnarray}

The complete secular equations of motion:
\begin{eqnarray}
\frac{dx_{{\rm S}}}{d\tau} & = &
\frac{5}{(1-e^{2}_{{\rm T}})^{\frac{3}{2}}}\sin^{2}{I}\frac{1-x_{{\rm
S}}^{2}}{(1-x_{{\rm S}}^{2}-y_{{\rm S}}^{2})^{\frac{1}{2}}}y_{{\rm S}}+\frac{5}{16}\alpha
\frac{e_{{\rm T}}}{(1-e^{2}_{{\rm T}})^{\frac{5}{2}}}(1-x_{{\rm S}}^{2}-\nonumber\\
& &
-y_{{\rm S}}^{2})^{\frac{1}{2}}[\sin{g_{{\rm T}}}\cos{I}(4+3(x_{{\rm
S}}^{2}+y_{{\rm S}}^{2})-5\sin^{2}{I}(1-x_{{\rm S}}^{2}+6y_{{\rm S}}^{2}))-\nonumber\\
& &
-10(1-x_{{\rm S}}^{2}-y_{{\rm S}}^{2})\sin{I}^{2}\cos{I}\sin{g_{{\rm
T}}}+2(3+5\sin^{2}{I})(y_{{\rm S}}^{2}\sin{g_{{\rm T}}}\times\nonumber\\
& & \times
\cos{I}+x_{{\rm S}}y_{{\rm S}}\cos{g_{{\rm
T}}})+20\sin^{2}{I}\cos{I}y_{{\rm S}}^{2}\sin{g_{{\rm
T}}}-70\sin^{2}{I}(y_{{\rm S}}^{2}\times\nonumber\\
& & \times
\sin{g_{{\rm T}}}\cos{I}+x_{{\rm S}}y_{{\rm S}}\cos{g_{{\rm
T}}})]+\frac{5}{16}\alpha\beta\frac{e_{{\rm T}}}{(1-e^{2}_{{\rm
T}})^{3}}[y_{{\rm S}}^{2}\sin{g_{{\rm T}}}(4+\nonumber\\
& &
+3(x_{{\rm S}}^{2}+y_{{\rm S}}^{2})-5\sin^{2}{I}(1-x_{{\rm
S}}^{2}+6y_{{\rm S}}^{2}))+10\cos{I}(y_{{\rm S}}^{2}\sin{g_{{\rm T}}}\times\nonumber\\
& &
\times\cos{I}+x_{{\rm S}}y_{{\rm S}}\cos{g_{{\rm T}}})(1-x_{{\rm
S}}^{2}+6y_{{\rm S}}^{2})+10(1-x_{{\rm S}}^{2}-y_{{\rm S}}^{2})y_{{\rm
S}}^{2}\times\nonumber\\
& & \times\sin{g_{{\rm T}}}(2\cos^{2}{I}-\sin^{2}{I})]+\frac{5}{16}\alpha
\frac{e_{{\rm T}}}{(1-e^{2}_{{\rm
T}})^{\frac{5}{2}}}\frac{1}{(1-x_{{\rm S}}^{2}-y_{{\rm S}}^{2})^{\frac{1}{2}}}\times\nonumber\\
& &
\times[y_{{\rm S}}^{2}\sin{g_{{\rm T}}}\cos{I}(4+3(x_{{\rm
S}}^{2}+y_{{\rm S}}^{2})-5\sin^{2}{I}(1-x_{{\rm S}}^{2}+6y_{{\rm S}}^{2}))+\nonumber\\
& &
+10(y_{{\rm S}}^{2}\sin{g_{{\rm T}}}\cos{I}+x_{{\rm S}}y_{{\rm
S}}\cos{g_{{\rm T}}})\cos^{2}{I}(1-x_{{\rm S}}^{2}+6y_{{\rm S}}^{2})+\nonumber\\
& & +10(1-x_{{\rm S}}^{2}
-y_{{\rm S}}^{2})y_{{\rm S}}^{2}\sin{g_{{\rm T}}}(2\cos^{2}{I}-\sin^{2}{I})\cos{I}]-\nonumber\\
& &
-\frac{\beta}{(1-e^{2}_{{\rm T}})^{2}}y_{{\rm S}}\cos{I}(1-x_{{\rm
S}}^{2}+4y_{{\rm S}}^{2})-\frac{1}{(1-e^{2}_{{\rm
T}})^{\frac{3}{2}}}y_{{\rm S}}\times\nonumber\\
& & \times
\frac{2-2x_{{\rm S}}^{2}+3y_{{\rm S}}^{2}}{(1-x_{{\rm S}}^{2}-y_{{\rm
S}}^{2})^{\frac{1}{2}}}\\
\frac{dy_{{\rm S}}}{d\tau} & = & -\frac{5}{(1-e^{2}_{{\rm
T}})^{\frac{3}{2}}}\sin^{2}{I}\frac{x_{{\rm S}}y_{{\rm
S}}^{2}}{(1-x_{{\rm S}}^{2}-y_{{\rm S}}^{2})^{\frac{1}{2}}}y_{{\rm S}}+\frac{5}{16}\alpha
\frac{e_{{\rm T}}}{(1-e^{2}_{{\rm T}})^{\frac{5}{2}}}(1-\nonumber\\
& &
-x_{{\rm S}}^{2}-y_{{\rm S}}^{2})^{\frac{1}{2}}[-\cos{g_{{\rm
T}}}(4+3(x_{{\rm S}}^{2}+y_{{\rm S}}^{2})-5\sin^{2}{I}(1-x_{{\rm S}}^{2}+\nonumber\\
& &
+6y_{{\rm S}}^{2}))-2(3+5\sin^{2}{I})(x_{{\rm S}}y_{{\rm
S}}\sin{g_{{\rm T}}}\cos{I}+x_{{\rm S}}^{2}\cos{g_{{\rm T}}})-\nonumber\\
& & -20\sin^{2}{I}\cos{I}x_{{\rm S}}y_{{\rm S}}\sin{g_{{\rm T}}}]+\frac{5}{16}\alpha
\beta \frac{e_{{\rm T}}}{(1-e^{2}_{{\rm T}})^{3}}[-x_{{\rm S}}y_{{\rm
S}}\sin{g_{{\rm T}}}\times\nonumber\\
& &
\times(4+3(x_{{\rm S}}^{2}+y_{{\rm S}}^{2})-5\sin^{2}{I}(1-x_{{\rm
S}}^{2}+6y_{{\rm S}}^{2}))-10\cos{I}(x_{{\rm S}}\times\nonumber\\
& & \times
y_{{\rm S}}\sin{g_{{\rm T}}}\cos{I}+x_{{\rm S}}^{2}\cos{g_{{\rm
T}}})(1-x_{{\rm S}}^{2}+6y_{{\rm S}}^{2})-10(1-x_{{\rm S}}^{2}-\nonumber\\
& &
-y_{{\rm S}}^{2})x_{{\rm S}}y_{{\rm S}}\sin{g_{{\rm T}}}(2\cos^{2}{I}-\sin^{2}{I})]+\frac{5}{16}\alpha
\frac{e_{{\rm T}}}{(1-e^{2}_{{\rm T}})^{\frac{5}{2}}}\times\nonumber\\
& & \times
\frac{1}{(1-x_{{\rm S}}^{2}-y_{{\rm S}}^{2})^{\frac{1}{2}}}[-x_{{\rm
S}}y_{{\rm S}}\sin{g_{{\rm T}}}\cos{I}(4+3(x_{{\rm S}}^{2}+y_{{\rm S}}^{2})-5\sin^{2}{I}\times\nonumber\\
& & \times
(1-x_{{\rm S}}^{2}+6y_{{\rm S}}^{2}))-10(x_{{\rm S}}y_{{\rm
S}}\sin{g_{{\rm T}}}\cos{I}+x_{{\rm S}}^{2}\cos{g_{{\rm T}}})\cos^{2}{I}(1-\nonumber\\
& &
-x_{{\rm S}}^{2}+6y_{{\rm S}}^{2})-10(1-x_{{\rm S}}^{2}-y_{{\rm
S}}^{2})x_{{\rm S}}y_{{\rm S}}\sin{g_{{\rm T}}}(2\cos^{2}{I}-\sin^{2}{I})\times\nonumber\\
& & \times \cos{I}]+\frac{\beta}{(1-e^{2}_{{\rm T}})^{2}}x_{{\rm
S}}\cos{I}(1-x_{{\rm S}}^{2}+4y_{{\rm S}}^{2})+\frac{1}{(1-e^{2}_{{\rm
T}})^{\frac{3}{2}}}x_{{\rm S}}\times\nonumber\\
& & \times \frac{2-2x_{{\rm S}}^{2}+3y_{{\rm S}}^{2}}{(1-x_{{\rm
S}}^{2}-y_{{\rm S}}^{2})^{\frac{1}{2}}}\\
\frac{dg_{{\rm T}}}{d\tau} & = &
\frac{1}{2}\frac{\beta}{(1-e^{2}_{{\rm T}})^{2}}[4+x_{{\rm
S}}^{2}+11y_{{\rm S}}^{2}-5\sin^{2}{I}(1-x_{{\rm S}}^{2}+4y_{{\rm S}}^{2})]+\nonumber\\
& & +\frac{1}{(1-x_{{\rm S}}^{2}-y_{{\rm
S}}^{2})^{\frac{1}{2}}(1-e^{2}_{{\rm
T}})^{\frac{3}{2}}}\cos{I}(1-x_{{\rm S}}^{2}+4y_{{\rm S}}^{2})-\frac{5}{16}\alpha\beta\times\nonumber\\
& & \times
\frac{1+4e^{2}_{{\rm T}}}{(1-e^{2}_{{\rm T}})^{3}e_{{\rm T}}}[(y_{{\rm
S}}\sin{g_{{\rm T}}}\cos{I}+x_{{\rm S}}\cos{g_{{\rm T}}})(4+3(x_{{\rm
S}}^{2}+y_{{\rm S}}^{2})-\nonumber\\
& &
-5\sin^{2}{I}(1-x_{{\rm S}}^{2}+6y_{{\rm S}}^{2}))-10(1-x_{{\rm
S}}^{2}-y_{{\rm S}}^{2})\sin^{2}{I}\cos{I}\times\nonumber\\
& & \times
y_{{\rm S}}\sin{g_{{\rm T}}}]+(\frac{5}{16}\alpha\frac{e_{{\rm
T}}}{(1-e^{2}_{{\rm T}})^{\frac{5}{2}}}\frac{1}{(1-x_{{\rm
S}}^{2}-y_{{\rm S}}^{2})^{\frac{1}{2}}}+\frac{5}{16}\alpha\beta\times\nonumber\\
& & \times
\frac{e_{{\rm T}}}{(1-e^{2}_{{\rm T}})^{3}}\cos{I})[-y_{{\rm
S}}\sin{g_{{\rm T}}}(4+3(x_{{\rm S}}^{2}+y_{{\rm S}}^{2})-5\sin^{2}{I}(1-\nonumber\\
& &
-x_{{\rm S}}^{2}+6y_{{\rm S}}^{2}))-10(y_{{\rm S}}\sin{g_{{\rm
T}}}\cos{I}+x_{{\rm S}}\cos{g_{{\rm T}}})\cos{I}(1-x_{{\rm S}}^{2}+\nonumber\\
& &
+6y_{{\rm S}}^{2})-10(1-x_{{\rm S}}^{2}-y_{{\rm S}}^{2})y_{{\rm
S}}\sin{g_{{\rm T}}}(2\cos^{2}{I}-\sin^{2}{I})]\\
\frac{de_{{\rm T}}}{d\tau} & = & \frac{5}{16}\frac{\alpha
\beta}{(1-e^{2}_{{\rm T}})^{2}}[(y_{{\rm S}}\cos{g_{{\rm
T}}}\cos{I}-x_{{\rm S}}\sin{g_{{\rm T}}})(4+3(x_{{\rm S}}^{2}+y_{{\rm S}}^{2})-\nonumber\\
& & -5\sin^{2}{I}(1-x_{{\rm S}}^{2}+6y_{{\rm S}}^{2}))-10(1-x_{{\rm
S}}^{2}-y_{{\rm S}}^{2})y_{{\rm S}}\sin^{2}{I}\times \nonumber\\
& & \times \cos{I}\cos{g_{{\rm T}}}]\\
\frac{dI}{d\tau} & = & -\frac{x_{{\rm S}}\dot{x}_{{\rm S}}+y_{{\rm
S}}\dot{y}_{{\rm S}}}{(1-x_{{\rm S}}^{2}-y_{{\rm
S}}^{2})^{\frac{1}{2}}}(\frac{1}{\tan{I}(1-x_{{\rm S}}^{2}-y_{{\rm
S}}^{2})^{\frac{1}{2}}}+\frac{\beta}{\sin{I}(1-e^{2}_{{\rm T}})^{\frac{1}{2}}})-\nonumber\\
& & -\frac{e_{{\rm T}}\dot{e}_{{\rm T}}}{(1-e^{2}_{{\rm
T}})^{\frac{1}{2}}}(\frac{1}{\beta \sin{I}(1-x_{{\rm S}}^{2}-y_{{\rm
S}}^{2})^{\frac{1}{2}}}+\frac{1}{\tan{I}(1-e^{2}_{{\rm T}})^{\frac{1}{2}}})
\end{eqnarray}
\begin{acknowledgements}
The author is grateful to Prof. Douglas Heggie for all the useful
discussions on the context of this paper. The author also thanks Seppo
Mikkola, who kindly provided the code for integrating hierarchical
triple systems.
\end{acknowledgements}

\end{article}
\end{document}